\documentclass[aps,prl,reprint,superscriptaddress]{revtex4-1}

\usepackage{graphicx}
\usepackage{physics}
\usepackage{stmaryrd,bm,amssymb,amsfonts,amsmath} % For \llbracket and \rrbracket
\usepackage{color}
\usepackage[colorlinks]{hyperref}

\newcommand{\mom}[1]{\left\langle{#1}\right\rangle}

\newcommand{\fref}[1]{Fig.~\ref{#1}}

\renewcommand{\section}[1]{{\smallskip\bf{#1}}.~}

\newcommand{\figext}{eps}
\newcommand{\figLoc}{./}

\newcommand{\K}{\ensuremath{\,\text{K}}}
\newcommand{\mK}{\ensuremath{\,\text{mK}}}
\newcommand{\GHz}{\ensuremath{\,\text{GHz}}}
\newcommand{\MHz}{\ensuremath{\,\text{MHz}}}
\newcommand{\Ohm}{\ensuremath{\,\Omega}}
\newcommand{\MSaps}{\ensuremath{\,\text{MSa/s}}}
\newcommand{\dB}{\ensuremath{\,\text{dB}}}

\newcommand{\calA}{\textsf{A}}
\newcommand{\calB}{\textsf{B}}
\newcommand{\calC}{\textsf{C}}
\newcommand{\calBp}{\textsf{B'}}

\newcommand{\UdeS}{\affiliation{Institut Quantique, D\'epartement de Physique, Universit\'e de Sherbrooke, Sherbrooke, Qu\'ebec J1K 2R1, Canada}}
\newcommand{\Poly}{\affiliation{D\'epartment de G\'enie Physique, Polytechnique Montr\'eal, Montr\'eal, Qu\'ebec H3T 1J4, Canada}}



\begin{document}

\title{Photocount statistics of the Josephson parametric amplifier: a question of detection}
    \author{Jean Olivier Simoneau} \email{Jean.Olivier.Simoneau@USherbrooke.ca} \UdeS
    \author{St\'{e}phane Virally} \UdeS \Poly
    \author{Christian Lupien} \UdeS
    \author{Bertrand Reulet} \UdeS
\date{\today}

\begin{abstract}
Parametric amplifiers are known to squeeze the vacuum state of the
electromagnetic field, which results in predictable statistics of the
photocounts at their output. However, recent theoretical work \cite{padurariu2012}
predicts a very different statistical distribution for an amplifier based on a
Josephson junction. We test the hypothesis experimentally and recover the expected
squeezed vacuum statistics.
We explain this discrepancy by showing theoretically how the photocount statistics is dictated by the detection process, from single mode (our experiment) to multimode, fully resolved in frequency (as in \cite{padurariu2012}). 
\end{abstract}

 % Need to verify/adjust the pertinence of PACS numbers

\maketitle

\section{Introduction}
Photons truly reveal themselves as particles only when interacting with the matter field. From the experimental perspective, a photon is best described as two causally linked events, a creation and an annihilation. The statistics of photocounts must then depend both on the emission and the detection modes, and predictions about statistics of photons emitted by any system should always specify the detection setup.

This fact becomes an important factor in some experiments. Consider, for instance, the output of a Josephson parametric amplifier (JPA) \cite{castellanos2009,bergeal2010,zhou2014,eichler2013, jebari2018}. 
This type of device is very much at the forefront of quantum optics in microwaves, as it constitutes a quantum-limited amplifier in this band and as such is likely to be used in all quantum computing and measurement schemes. Understanding the noise characteristics of those devices is critical for these typically small signal applications. In the most fundamental case, the input of a JPA is simply the electromagnetic vacuum. A theoretical paper~\cite{padurariu2012} predicts the full counting statistics of photocounts emitted by such a system. But the results appear to contradict the model of parametric amplifiers as ``vacuum squeezers'', a well-studied quantum optics fact. The squeezing operator generates pairs of photons, and this is reflected in the photocount variance, which reads $\mom{\delta n^2}=2\mom{n}\left(\mom{n}+1\right)$. In contrast, the theoretical predictions of Ref.~\cite{padurariu2012} is $\mom{\delta n^2}=2\mom{n}\left(8\mom{n}^2+5\mom{n}+1\right)$\footnote[2]{For a specific value of the photocount integration time, namely the inverse of the JPA's cavity bandwidth.}\cite{Vyas1989,DeBrito1996}. In both cases, at very small signal, $\mom{\delta n^2}\simeq2\mom{n}$, twice the classical value. This reflects the emission of pairs of photons in the squeezing process. However, the variance predicted in Ref.~\cite{padurariu2012} dramatically increases as $\mom{n}^3$ for higher signals. This is a strong departure form the expected squeezed vacuum behavior.

In this paper, we show that the apparent discrepancy is due to the choice of detection scheme in the theoretical reference. Indeed, the detector is assumed to have infinite frequency resolution. Of course, real measurements are limited both in time and bandwidth. They inherently possess finite frequency resolution. When the detection bandwidth closely resembles the natural mode of the amplifier's cavity, a single quantum mode is observed and we show both theoretically and experimentally that the `correct' squeezed vacuum statistics is recovered.

The paper is organized as follows. We present an experiment with limited frequency resolution at the output of a JPA with vacuum input. The discrete photocount statistics is recovered from continuous voltage measurements~\cite{virally2016}. We show that after careful calibration, we recover a variance and third-order photocount moment equal to those predicted for a squeezed vacuum, and not those predicted by Ref.~\cite{padurariu2012}. These measurements are well captured by a simple input-output~\cite{Gardiner1985,Collett1987,Gardiner2004}
model of the JPA, followed by a single-mode (non frequency-resolved) detector. In contrast, a variant of the model, using the same input-output relations for the JPA but a multimode (frequency-resolved) detector, leads to the statistics predicted by Ref.~\cite{padurariu2012}. The distinction is lost in narrow-band experiments, but we anticipate that it will play a crucial role in the results of future experiments using the new generation of wide bandwidth JPAs~\cite{macklin2015,mendes2019}.

\section{Experimental setup}
\begin{figure}
    \centering
    \includegraphics[width=\columnwidth]{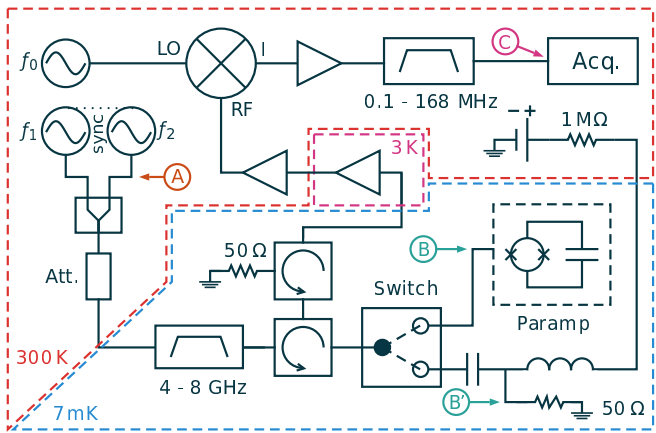}
    \caption{
        Experimental setup used for detection.
        %The noise of the paramp is downconverted towards dc using an IQ mixer. 
        %It is then filtered and digitized. 
        The flux bias coil of the paramp is omitted for clarity. 
        %The $50\,\Omega$ resistor is used for calibration purposes. 
        Circled letters are calibration reference points. See text for details.
    }
    \label{fig_Setup}
\end{figure}

The experimental setup is presented in \fref{fig_Setup}. 
We study the signal emitted by a commercial Josephson parametric amplifier (paramp), similar to that of Ref. \cite{hatridge2011}, placed in a dilution refrigerator at $\sim7\mK$ and  driven by two (phase-locked) sinusoidal pumps of frequencies $f_1=4.5\GHz$ and $f_2=7.5\GHz$. 
The output signal is  measured in a small frequency band centered around $(f_1+f_2)/2=6.0\GHz$. 
The dual-pump operation mode \cite{kamal2009} is selected to avoid residual pump signal in the measurement band, so that the input of the paramp in the measured bandwidth can be considered as the vacuum.

The paramp resonance frequency can be tuned by a current bias through a superconducting flux coil in the vicinity of the paramp SQUID loop (omitted on schematic for clarity). 
A 4--8 GHz band pass filter protects the paramp from radiation outside of its operation range.
Circulators are used to separate the input and output fields of the paramp and to isolate it from the noise of the $3\K$ and $300\K$ stages. 
A microwave switch is used to swap a $50\Ohm$ resistor in place of the paramp for calibration purposes.

The paramp output signal is amplified and conveyed to $300\K$, where it is downconverted by an IQ mixer with a local oscillator (LO) at frequency $f_0\approx6.0\GHz$. 
The LO is \emph{not} phase-locked with the pumps. 
The downconverted signal is then filtered by a 0.1 -- $168\MHz$ bandpass filter and sampled by a fast acquisition card with 14-bit resolution and $400\MSaps$ rate. The effective photocount integration time is the inverse of the full detected bandwidth ($2\times168\MHz$).
Histograms of the measured signal and their six first cumulants are computed on the fly during the data acquisition.

\section{Calibration}
Proper calibration is essential to compare experimental results with theory. Three calibrations are required: that of the ac power at the sample level, that of the absolute photon numbers that are detected, calculated for the measured voltage cumulants $C_j$, and that of the paramp resonance frequency vs. current in the flux bias coil.  

To calibrate the attenuation between the excitation at room temperature (circle reference \calA\ in \fref{fig_Setup}) and the input port of the paramp ($\calB\approx\calBp$), we use a macroscopic $R = 50\Ohm$ resistor in place of the paramp (using the cryogenic switch). We can heat that resistor using either a known dc current or an ac bias and observe the temperature increase by the increased noise it emits. Thus we can map which dc current is needed to heat the resistor as much as a given ac voltage, as in \cite{santavica2010}. The linear relation we observe between them provides us with the  \calA--\calB\ attenuation, $24.96\dB$.

To calibrate the effective gain between the output of the paramp (\calB) and the data acquisition (\calC), we measure the \calA--\calC\ gain by adjusting the paramp DC flux line to put it out of resonance such that it totally reflects an incoming test tone signal of known amplitude, and subtract the previously obtained \calA--\calB\ attenuation. We find the \calB--\calC\ gain to be $87.67\dB$. 

The paramp resonance frequency, which is controlled by the current applied to the flux bias coil, is calibrated by measuring the reflected phase on the paramp using a vector network analyser in the absence of a pump signal.\cite{hatridge2011, mutus2013}

\section{Measurements}
\begin{figure}
    \centering
    \includegraphics[width=\columnwidth]{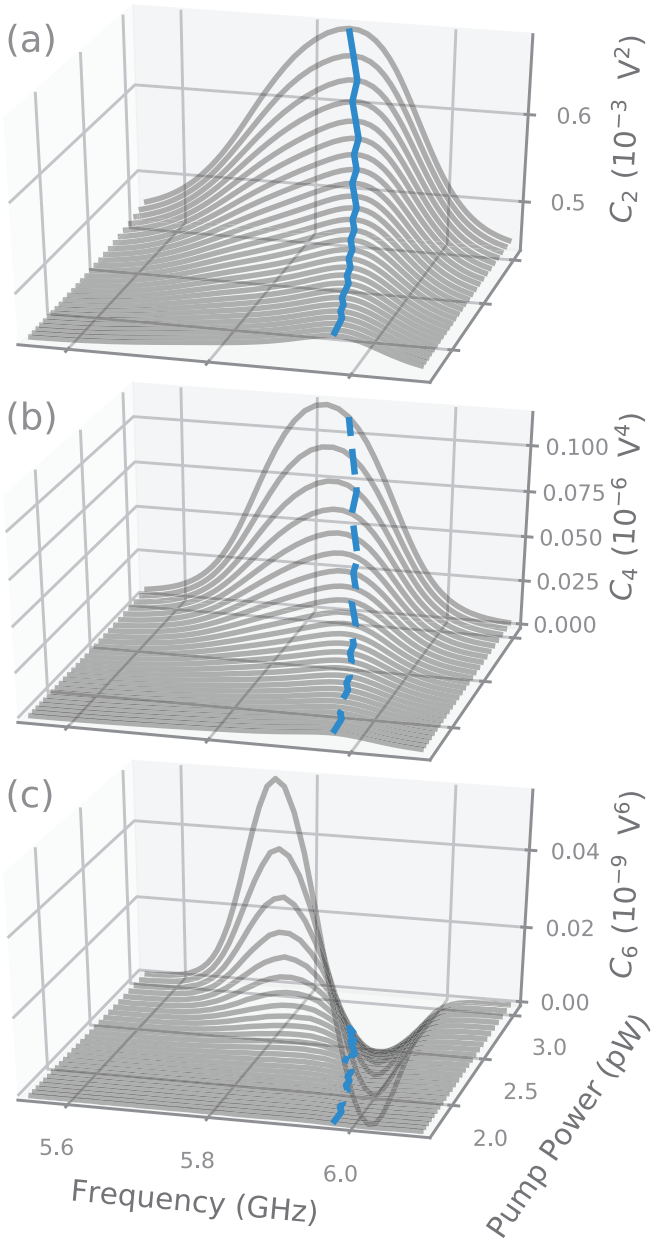}
    \caption{Measured cumulants of voltage fluctuations generated by the paramp as a function of its resonance frequency and pump power. 
    The \emph{Frequency} axis is controlled by the flux bias. The thick blue solid line in (a) corresponds to the ridge of $C_2$. The dashed blue line in (b) and (c) corresponds to frequency and pump power that correspond to the blue line of (a).  
    }
    \label{fig_Cumulants}
\end{figure}

\begin{figure}
    \centering
    \includegraphics[width=\columnwidth]{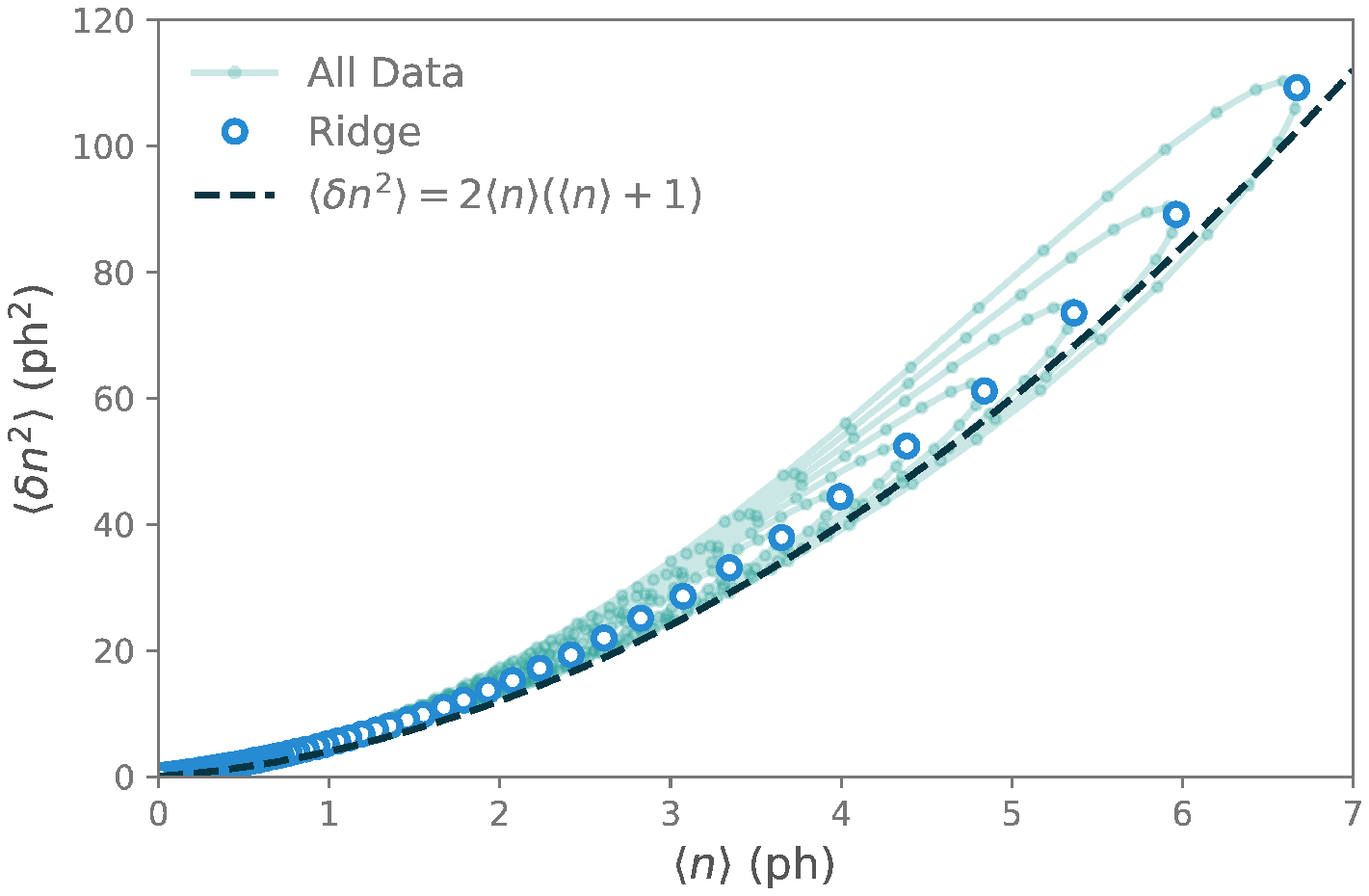}
    \caption{Variance of the photocounts $\langle\delta n^2\rangle$ as a function of the average photon number $\langle n\rangle$. 
    %The light circles are experimental points. 
    %The dark circles correspond to the maximum of the paramp gain, i.e the blue line in \fref{fig_Cumulants}. 
    Dots are experimental data and each line represents a given pump power.
    Open circles correspond to the maximum of the paramp gain, i.e the blue line in \fref{fig_Cumulants}.
    The dashed line corresponds to the theoretical prediction for squeezed vacuum. }
    \label{fig_PhotonStats2}
\end{figure}

\begin{figure}
    \centering
    \includegraphics[width=\columnwidth]{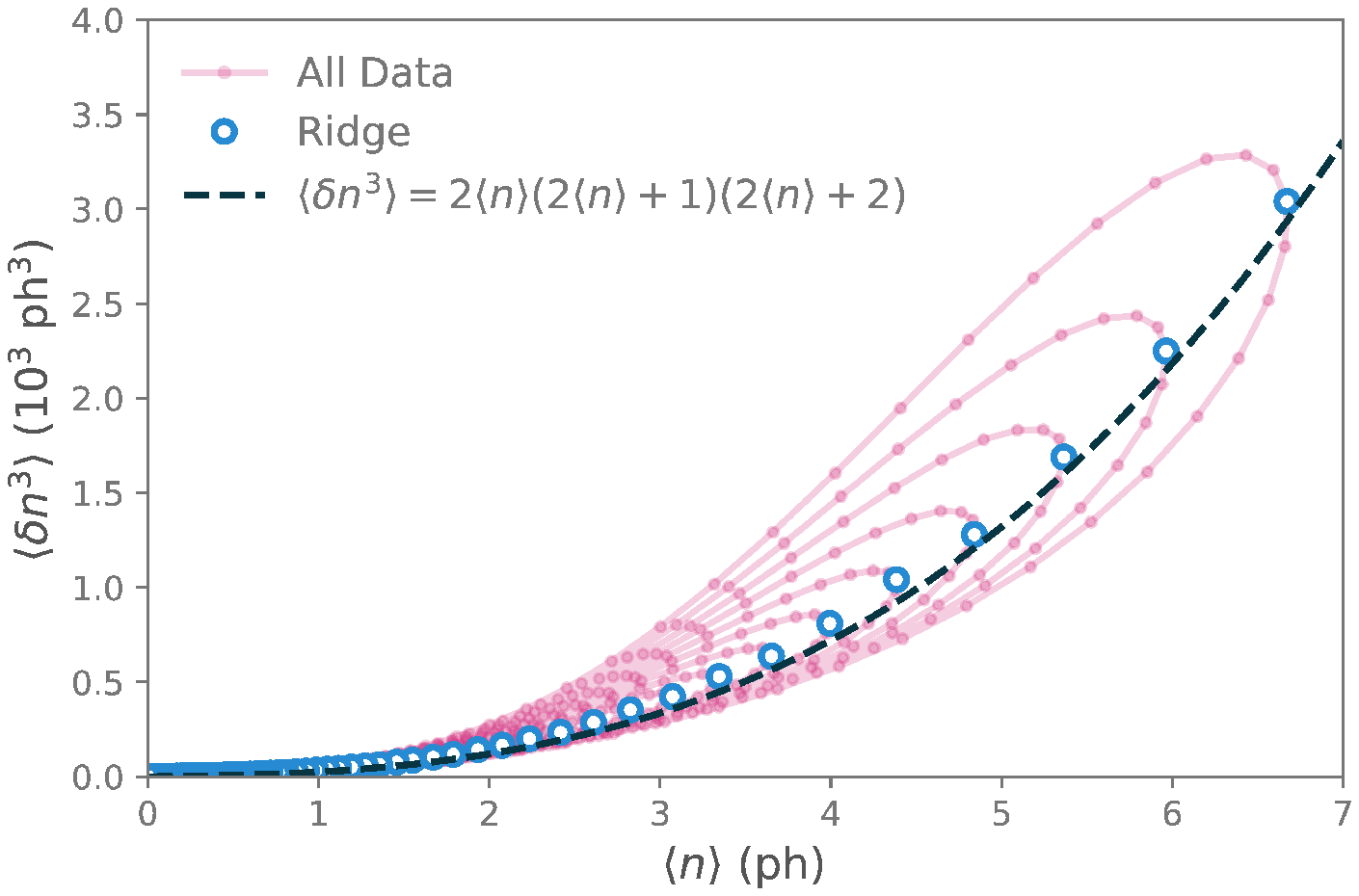}
    \caption{Skewness of the photocounts $\langle\delta n^3\rangle$ as a function of the average photon number $\langle n\rangle$. 
    %The light circles are experimental points. 
    %The dark circles correspond to the maximum of the paramp gain, i.e the blue line in \fref{fig_Cumulants}. 
    Dots are experimental data and each line represents a given pump power.
    Open circles correspond to the maximum of the paramp gain, i.e the blue line in \fref{fig_Cumulants}.
    The dashed line corresponds to the theoretical prediction for squeezed vacuum. }
    \label{fig_PhotonStats3}
\end{figure}

In order to probe the photon statistics of the paramp for different regimes of operations, we explore its parameter space, flux bias and pump power, for a fixed measurement frequency $f_0 = (f_1+f_2)/2 = 6.0\GHz$.
Experimentally, we first select a pump power yielding a maximum gain of approximately $10\dB$ and adjust the paramp at this operation point.
Then, we sweep the flux bias current and the pump power around the initial values while measuring the cumulants of the voltage fluctuations generated by the paramp. From these we compute the moments of the photocount distribution $\langle n\rangle, \langle \delta n^2\rangle, \langle \delta n^3\rangle$, shown in \fref{fig_Cumulants} using the precedure developped in \cite{virally2016, simoneau2017}. We show in \fref{fig_PhotonStats2} the variance and in \fref{fig_PhotonStats3} the skwness of the photocount distribution as a function of the average photon number. There are many combinations of flux bias and pump power that give the same average photon number $\langle n\rangle$, each providing a different value of $\langle \delta n^2\rangle$ and $\langle \delta n^3\rangle$. As a consequence, Figs. \ref{fig_PhotonStats2} and \ref{fig_PhotonStats3} exhibit clouds of experimental points and not just single curves. A particular subset of points corresponds to the maximum gain of the paramp, i.e. the largest value of $\langle n\rangle$ for each pump power. Those are the best operating points for the paramp used as an amplifier; they are represented as blue solid points in \fref{fig_Cumulants} (a). Reporting these points in Figs. \ref{fig_Cumulants} (b) and (c), we observe that they are close the the maximum of the fourth cumulant $C_4$ and correspond to a vanishing $C_6$. The same points are highlighted in Figs. \ref{fig_PhotonStats2} and \ref{fig_PhotonStats3} (open circles). We find that these specific points closely follow the expected relations for a squeezed vacuum, represented by dashed lines in Figs. \ref{fig_PhotonStats2} and \ref{fig_PhotonStats3}.


\section{Theory}
To model the JPA, we apply the input-output formalism~\cite{Gardiner1985,Collett1987,Gardiner2004}
to a single-ended, frequency-symmetric single-mode cavity.
The intra-cavity Hamiltonian is assumed to be the squeezing Hamiltonian that
is characteristic of parametric amplifiers~\cite{Mollow1967b,Mollow1967c,Collett1987a,Boutin2017}. 
The output electromagnetic modes can be written as a function of the free input modes $\bm{b}_\nu$, in the frame rotating at $\nu_0$ and up to a constant phase, as
\begin{equation}
    \bm{B}_\textrm{out}(\nu)=\cosh[\eta(\nu)]\bm{b}_\nu+e^{i\phi}\sinh[\eta(\nu)]\bm{b}^\dagger_{-\nu},
\end{equation}
with $\phi$ defining a squeezing direction, and
\begin{equation}
  \label{eta}
  \eta(\nu)=\frac{1}{2}\ln\!\left[
    \frac{\sqrt{(\Gamma^2+\nu^2+\left\lvert\xi\right\rvert^2-\delta^2)^2
        +4\delta^2\Gamma^2}+2\Gamma\left\lvert\xi\right\rvert}
    {\sqrt{(\Gamma^2+\nu^2+\left\lvert\xi\right\rvert^2-\delta^2)^2+4\delta^2\Gamma^2}
      -2\Gamma\left\lvert\xi\right\rvert}\right].
\end{equation}
In this expression, $\Gamma$ is the cavity coupling parameter, inversely proportional
to the decay time of the cavity, and $\xi$ is the nonlinear intra-cavity 2-photon coupling parameter (in the two-pump scheme, $\lvert\xi\rvert\propto\sqrt{P_1\,P_2}$, the geometric average of both pump powers). The photon-photon interaction Hamiltonian also shifts the
position of the center peak of the cavity mode proportionally to $P_1+P_2$, as
seen in \fref{fig_Cumulants}. The peak can be brought back to the center of the
measurement window by adjusting the magnetic flux. This is captured by
$\delta=\phi+\lvert\xi\rvert(P_1+P_2) /\sqrt{P_1\,P_2}$, where $\phi$ is the frequency
shift induced by the magnetic flux. The `ridge' (maximum) of $C_2$ observed in
\fref{fig_Cumulants} corresponds to $\delta=0$.

\begin{figure}[t]
    \centering
    \includegraphics[width=0.51\textwidth]{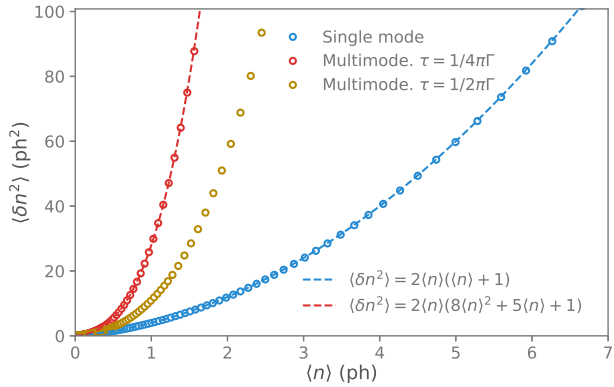}
    \caption{Comparison between single mode counting, as in our experiment, and multimode counting with unit cavity coupling strength, as in Ref.~\cite{padurariu2012}. The parameter $\tau$ is the time of accumulation of photocounts for each data point~\cite{Note1}.}
    \label{fig_usvsNaz}
\end{figure}

Using this model, we calculate the moments of the statistics of the photon flux
per unit of frequency and time, for a detector with normalized response function $h(\nu)$,
\begin{multline}
    \label{eq=us}
    \mom{n^k}=\int \dd{\nu_1}\cdots \dd{\nu_{2k}}\,
    h^*(\nu_1)\,h(\nu_2)\cdots h^*(\nu_{2k-1})\,h(\nu_{2k})\\
    \bra{\phi_\mathrm{i}}\bm{B}_\mathrm{out}^\dagger(\nu_1)\bm{B}_\mathrm{out}(\nu_2)\cdots
    \bm{B}_\mathrm{out}^\dagger(\nu_{2k-1})\bm{B}_\mathrm{out}(\nu_{2k})\ket{\phi_\mathrm{i}},
  \end{multline}
where $\ket{\phi_\mathrm{i}}$ is the input state of the JPA. In the case of an
electromagnetic vacuum input, $\ket{\phi_\mathrm{i}}=\ket{\textrm{vac}}$, we find the expected
statistics of a squeezed vacuum on the `ridge' of $C_2$. In particular, for $\eta$ in the unit range or above,~\footnote[1]{See supplementary material.}
\begin{equation}
    \mom{n}=\int \dd{\nu}\left\lvert h(\nu)\right\rvert^2n(\nu),
\end{equation}
with $n(\nu)=\sinh^2[\eta(\nu)]$, and
\begin{equation}
    \label{res=us}
        \mom{\delta n^2}=2\mom{n}(\mom{n}+1).
\end{equation}
The Fano factor $\mathcal{F}=\mom{\delta n^2} /\mom{n}$ is thus simply
\begin{equation}
    \label{Fano_sqvac}
    \mathcal{F}=2\,(\mom{n}+1).
\end{equation}

This result is at odds with
the predictions of Ref.~\cite{padurariu2012}. The reason is that the theoretical framework of the reference uses a different detection scheme, where the signal is resolved in frequency. For a frequency resolution $\Delta$, the measured moments per unit time are,
\begin{multline}
    \label{eq=naz}
    \mom{n^k}_\Delta=\int\dd{\nu_1}\cdots\dd{\nu_{2k}}\,
    \delta_\Delta(\nu_1-\nu_2)\cdots\delta_\Delta(\nu_{2k-1}-\nu_{2k})\\
    \bra{\phi_\mathrm{i}}\bm{b}_\mathrm{o}^\dagger(\nu_1)\bm{b}_\mathrm{o}(\nu_2)\cdots
     \bm{b}_\mathrm{o}^\dagger(\nu_{2k-1})\bm{b}_\mathrm{o}(\nu_{2k})\ket{\phi_\mathrm{i}},
\end{multline}
where $\delta_\Delta$ are peaked functions of width $\Delta$ and unit integrated value. They tend to the true Dirac delta distribution as $\Delta\to 0$~\cite{Note1}. In the same limit, the Fano factor $\mathcal{F}_\Delta\equiv\mom{\delta n^2}_\Delta /\mom{n}_\Delta$ behaves as
\begin{equation}
    \label{res=naz}
    \lim_{\Delta \to 0} \mathcal{F}_\Delta=\frac{\int \dd{\nu}2\,n(\nu)[n(\nu)+1]}{\int \dd{\nu}n(\nu)}.
\end{equation}

This result is very different from that of Eq.~\eqref{res=us}. It corresponds to summing many independent modes, resolved in frequency. Each mode is a squeezed vacuum with a Fano factor of the form of Eq.~\eqref{Fano_sqvac}. But the behavior of the overall Fano factor is dependent on the detector bandwidth (the inverse of the time $\tau$ spent accumulating photocounts for each data point~\cite{Note1}). For a detector with bandwidth $4\pi\Gamma$ (the coupling strength between the inside and outside of the cavity), the Fano factor corresponds to that found in Ref.~\cite{padurariu2012}. However, if another bandwidth is chosen, the relation between $\mom{\delta n^2}$ and $\mom{n}$ is different, as shown in \fref{fig_usvsNaz}.

\section{Discussion}
Our theoretical analysis clearly explains the importance of the detection scheme and thus why the predictions of \cite{padurariu2012} differ from that of the expected squeezed vacuum statistics. It also predicts that with our present setup we should observe the photon statistics of squeezed vacuum. We indeed observe the `right' statistics on the `ridge' of $C_2$, that is on the optimal functioning points of the paramp.

However, the theory fails to explain why we observe clouds of points for $\langle\delta n^2\rangle$ and $\langle\delta n^3\rangle$ vs. $\langle n\rangle$ in Figs. \ref{fig_PhotonStats2} and \ref{fig_PhotonStats3}. It does predict that we can observe a photocount variance lower than that of the squeezed vacuum, but never higher~\cite{Note1}. One can easily understand that if the measurement bandwidth is finite and not centered on the resonance (i.e., off the ridge) there might be photon pairs which are detected as single photons (the other photon of the pair being outside the detection bandwidth). This leads to a mixture between squeezed vacuum and thermal state and thus leads to a decrease of $\langle\delta n^2\rangle$.
In contrast, we observe that experimental points off the ridge lie both below and above the variance of squeezed vacuum.

In an attempt to correct the theory, we considered the effect of wideband detection~\cite{Note1, Virally2019}. We do find a small cloud of points, but they all lie beneath the theoretical maximum variance. In addition, we find a non-zero sixth-cumulant $C_6$ outside of the ridge, as featured in Fig.~\ref{fig_Cumulants}. This is an interesting feature, as the narrow-band theory predicts $C_6=0$ everywhere, just as it is zero on the `ridge' of our experiments. However, the amplitude of the corrected $C_6$ is too small by one order of magnitude compared to the experiments. Hence, the wideband correction is insufficient to explain experimental data.

Another potential shortcoming of the theoretical model is the fact that the nonlinear coupling of the Josephson junction is cut at the second order in our Hamiltonian. This is also the case for the reference motivating this text~\cite{padurariu2012}, so we did not attempt to expand the Hamiltonian to higher orders. However, the Josephson parametric amplifier can be highly nonlinear, and this simplification is likely to fail at higher powers, starting at the single digit photon number. This has been studied in details in \cite{Boutin2017} for the case $\phi=0$. In Figs.~\ref{fig_PhotonStats2} and~\ref{fig_PhotonStats3}, we linked all the points corresponding to the same pump power by a single line. We clearly see that excursions away from the theoretical values increase dramatically with pump power. We also see that the ridge is defined as the maximum of $\langle n\rangle$ vs. flux bias for any given pump power, i.e. $\left(\frac{\partial\langle n\rangle}{\partial \Phi}\right)_P=0$. It is straightforward to show that it also corresponds to the minimum pump power for a given $\langle n\rangle$, i.e. $\left(\frac{\partial P}{\partial \Phi}\right)_{\langle n\rangle}=0$. As a consequence, we show that we recover the squeezed vacuum photocount distribution only for a pump power close to the optimum gain (even though our bichromatic pumping scheme is the one that leads to the least nonlinearities \cite{Boutin2017}). In addition, half the points lie above, and half the points lie below the theoretical curve. Thus we expect that a successful theory would take into account the sign of the flux bias (i.e. it should feature odd terms in the flux bias, which our theory fails to do).

\section{Conclusion}
We have performed an experimental and theoretical investigation of the photon statistics of the microwave radiation generated by a Josephson parametric amplifier. We have observed that with a wideband, single-mode detection scheme, the statistics is that of squeezed vacuum when the pump of the paramp is kept to its lowest value for a given average photon number and strongly departs from it at higher power. Our theoretical analysis shows how the photocount statistics crucially depends on the detection bandwidth, from a time-resolved, wideband amplifier (our setup) to that of frequency-resolved photodetection\cite{padurariu2012}. Our results, which are valid for any kind of paramp, are of great interest both to the development of quantum limited amplifiers with optimal photon statistics as well as for the development of sources of radiation with non-classical statistics. As a matter of fact, instead of playing with the source, we show that one can play with the detector, in a similar way as quantum computation may require non-Gaussian states of light if measurements are performed with linear detectors whereas Gaussian states are enough if one uses single photon detectors\cite{knill2001}. More theoretical and experimental works are needed to explore the path we have paved, in particular to understand how high order terms in the Hamiltonian affect the photcount distribution for an arbitrary detection bandwidth.

\vspace{\baselineskip}
\begin{acknowledgments}
We thank G. Laliberté for technical help and S. Boutin for fruitful discussions. This work was supported by the Canada Excellence Research Chair program, the NSERC, the Canada First Research Excellence Fund, the MDEIE, the FRQNT, the INTRIQ, the Université de Sherbrooke, and the Canada Foundation for Innovation.
\end{acknowledgments}

\bibliography{article}

\end{document}

\title{Photocount statistics of the Josephson parametric amplifier: a question of detection}
\author{Jean Olivier Simoneau}
\affiliation{Institut Quantique, D\'{e}partement de Physique, Universit\'{e} de Sherbrooke, Sherbrooke, Qu\'{e}bec J1K 2R1, Canada}
\author{St\'{e}phane Virally}
\affiliation{Institut Quantique, D\'{e}partement de Physique, Universit\'{e} de Sherbrooke, Sherbrooke, Qu\'{e}bec J1K 2R1, Canada}
\affiliation{femtoQ Lab, D\'{e}partement de G\'{e}nie Physique, Polytechnique
  Montr\'{e}al, Montr\'{e}al, Qu\'{e}bec H3T 1JK, Canada}
\author{Christian Lupien}
\affiliation{Institut Quantique, D\'{e}partement de Physique, Universit\'{e} de Sherbrooke, Sherbrooke, Qu\'{e}bec J1K 2R1, Canada}
\author{Bertrand Reulet}
\affiliation{Institut Quantique, D\'{e}partement de Physique, Universit\'{e} de Sherbrooke, Sherbrooke, Qu\'{e}bec J1K 2R1, Canada}
\date{\today}

\maketitle

We consider a single-sided, narrow-band cavity mode $\bm{A}(t)$ coupled on one side to a continuum of modes $\bm{B}_\nu(t)$. The coupling coefficient $\Gamma$ is assumed to be constant over the narrow range of frequencies under consideration.

The resonance of the cavity can be modified via a magnetic flux $\phi$. The cavity is pumped by two narrow-band signals with power $P_1$ and $P_2$ and respective frequencies $\nu_1$ and $\nu_2$. We detect the output signal in a narrow band around $\nu_0=(\nu_1+\nu_2)/2$.

We use the input-output formalism~\cite{Gardiner1985} to model the interaction between the cavity mode and the output modes~\cite{Boutin2017}.

\section{Input-output model}
\subsection{Evolution of the external operators}
The Hamiltonian for the external modes is
\begin{equation}
  \bm{H}_b=\int_{-\infty}^{+\infty}\dd{\nu}2\pi\hbar\nu\;\bm{B}^\dagger_\nu\bm{B}_\nu,
\end{equation}
while the interaction Hamiltonian is
\begin{equation}
  \label{Hi}
  \bm{H}_i=i\hbar \sqrt{4\pi\Gamma}\int_{-\infty}^{+\infty}\dd{\nu}\left[\bm{B}_\nu^\dagger\bm{A}-\bm{B}_\nu\bm{A}^\dagger\right].
\end{equation}

Strictly speaking, $\nu$ only goes from 0 to $+\infty$. However, as usual in quantum optics~\cite{Gardiner1985} in the narrow-band regime, i.e. when the modes are centered around a frequency $\nu_0$ and in a bandwidth $B\ll\nu_0$, we extend the limits of all integrations in $\nu$ from $-\infty$ to $+\infty$.

The $\bm{B}_\nu$ operators evolve as
\begin{equation}
    \dv{\bm{B}_\nu}{t}=\frac{i}{\hbar}\left[\bm{H}_b+\bm{H}_i,\bm{B}_\nu\right],
\end{equation}
and we find
\begin{widetext}
\begin{equation}
  \begin{split}
    \bm{B}_\nu(t)&=\bm{B}_\nu\!\left(t_\mathrm{in}\right)e^{-i2\pi\nu\left(t-t^\mathrm{in}\right)}+\sqrt{4\pi\Gamma}\int_{t^\mathrm{in}}^t\dd{\tau}\bm{A}(\tau)e^{-i2\pi\nu(t-\tau)},\\
    &=\bm{B}_\nu\!\left(t_\mathrm{out}\right)e^{-i2\pi\nu\left(t-t^\mathrm{out}\right)}-\sqrt{4\pi\Gamma}\int_{t}^{t^\mathrm{out}}\dd{\tau}\bm{A}(\tau)e^{-i2\pi\nu(t-\tau)},
  \end{split}
\end{equation}
where $t_\mathrm{in}$ and $t_\mathrm{out}$ are times before and after
interaction of the external modes with the cavity.

We define the integrated external modes
\begin{equation}
  \begin{split}
  &\bm{B}(t)=\int_{-\infty}^{+\infty}\dd{\nu}\bm{B}_\nu(t);\\
  &\bm{B}_\mathrm{in}(t)\equiv-\int_{-\infty}^{+\infty}\dd{\nu}\bm{B}_\nu\!\left(t_\mathrm{in}\right)e^{-i2\pi\nu(t-t_\mathrm{in})}\equiv\int_{-\infty}^{+\infty}\dd{\nu}\bm{B}_\mathrm{in}(\nu)\,e^{-i2\pi\nu t};\\
  &\bm{B}_\mathrm{out}(t)\equiv+\int_{-\infty}^{+\infty}\dd{\nu}\bm{B}_\nu\!\left(t_\mathrm{out}\right)e^{-i2\pi\nu(t-t_\mathrm{out})}\equiv\int_{-\infty}^{+\infty}\dd{\nu}\bm{B}_\mathrm{out}(\nu)\,e^{-i2\pi\nu t},
  \end{split}
\end{equation}
where the $\pm$ signs are only a convention. This leads to the input-output relations between the external modes before and after interaction,
\begin{equation}
  \bm{B}(t)=-\bm{B}_\mathrm{in}(t)+\frac{\sqrt{4\pi\Gamma}}{2}\bm{A}(t)=\bm{B}_\mathrm{out}(t)-\frac{\sqrt{4\pi\Gamma}}{2}\bm{A}(t).
\end{equation}
\end{widetext}

We assume that the input modes have not interacted with anything. That is, we identify
\begin{equation}
  \bm{B}_\mathrm{in}(\nu)=\bm{b}_\nu,
\end{equation}
where the $\bm{b}_\nu$ are the free modes of the EM field.

\subsection{Evolution of the cavity operator}
\subsubsection{Bare cavity}
The Hamiltonian for the internal mode is $\bm{H}_a$. It evolves as
\begin{equation}
  \dv{\bm{A}}{t}=\frac{i}{\hbar}\left[\bm{H}_a+\bm{H}_i,\bm{A}\right].
\end{equation}

For a bare, narrow-band cavity
\begin{equation}
  \bm{H}_a=2\pi\hbar\nu_0\;\bm{A}^\dagger\bm{A},
\end{equation}
leading to
\begin{equation}
  \begin{split}
    \dv{\bm{A}}{t}&=-i2\pi(\nu_0-i\Gamma)\bm{A}+\sqrt{4\pi\Gamma}\bm{B}_\mathrm{in};\\
    &=-i2\pi(\nu_0+i\Gamma)\bm{A}-\sqrt{4\pi\Gamma}\bm{B}_\mathrm{out}.
  \end{split}
\end{equation}

We first eliminate $\nu_0$ by placing ourselves in the rotating frame. In addition, we define the Fourier transform of $\bm{A}(t)$,
\begin{equation}
  \bm{A}(\nu)=\int_{-\infty}^{+\infty}\dd{t}\bm{A}(t)\;e^{i2\pi\nu t},
\end{equation}
and obtain
\begin{align}
    &-i2\pi(\nu+i\Gamma)\bm{A}(\nu)=\sqrt{4\pi\Gamma}\;\bm{b}_\nu;\\
    &-i2\pi(\nu-i\Gamma)\bm{A}(\nu)=-\sqrt{4\pi\Gamma}\;\bm{B}_\mathrm{out}(\nu),
\end{align}
Note that we have replaced $\bm{B}_\mathrm{in}(\nu)$ by
$\bm{b}_\nu$ in the first equality.

For a cavity with an infinite Q-factor, we would expect
\begin{equation}
  \bm{A}(\nu)=e^{i\phi(\nu)}\sqrt{\delta(\nu)}\;\bm{b}_\nu.
\end{equation}

Indeed, we get
\begin{widetext}
\begin{equation}
  \bm{A}(\nu)=\frac{\sqrt{\Gamma/\pi}}{\Gamma-i\nu}\;\bm{b}_\nu=e^{i\arctan(\nu/\Gamma)}\sqrt{\frac{\Gamma/\pi}{\Gamma^2+\nu^2}}\;\bm{b}_\nu,
\end{equation}
\end{widetext}
so that for $\Gamma\rightarrow\infty$, the expression indeed tends towards the square-root of the delta distribution.

We then find
\begin{equation}
  \bm{B}_\mathrm{out}(\nu)=e^{-i2\arctan(\nu/\Gamma)}\;\bm{b}_\nu,
\end{equation}
which shows that the output modes remain the free modes of the EM field, with a frequency-dependent phase, as expected. The input-output relation is unitary.

\subsubsection{Parametric amplifier}
For a narrow-band parametric amplifier, we have
\begin{equation}
  \bm{H}_a=2\pi\hbar\left(\nu_0+\delta\right)\bm{A}^\dagger\bm{A}+i2\pi\hbar\left(\xi\bm{A}^\dagger\bm{A}^\dagger-\xi^*\bm{A}\bm{A}\right),
\end{equation}
where $\xi$ is the nonlinear coefficient of the amplification process and
$\delta=\phi+\left\lvert\xi\right\rvert(P_1+P_2)/\sqrt{P_1P_2}$ ($\phi$ is
the frequency displacement provided by the magnetic flux).

We eliminate $\nu_0$ by placing ourselves in the rotating frame, and get
\begin{equation}
  \label{haA}
  \begin{split}
    &\frac{i}{\hbar}\left[\bm{H}_a,\bm{A}(t)\right]=-i2\pi\hbar\delta\bm{A}(t)+2\pi\xi\bm{A}^\dagger(t)\textrm{, or}\\
    &\frac{i}{\hbar}\left[\bm{H}_a,\bm{A}(\nu)\right]=-i2\pi\delta\bm{A}(\nu)+2\pi\xi\bm{A}^\dagger(-\nu),
  \end{split}
\end{equation}
as
\begin{equation}
  \begin{split}
    \bm{A}^\dagger(t)&=\int_{-\infty}^{+\infty}\dd{\nu}\bm{A}^\dagger(\nu)\;e^{i2\pi\nu t}\\&=\int_{-\infty}^{+\infty}\dd{\nu}\bm{A}^\dagger(-\nu)\;e^{-i2\pi\nu t}.
  \end{split}
\end{equation}

We then have to solve
\begin{widetext}
\begin{align}
    &-i2\pi(\nu+i\Gamma)\bm{A}(\nu)=-i2\pi\delta\bm{A}(\nu)+2\pi\xi\bm{A}^\dagger(-\nu)+\sqrt{4\pi\Gamma}\bm{b}_\nu;\label{eq1}\\
    &-i2\pi(\nu-i\Gamma)\bm{A}(\nu)=-i2\pi\delta\bm{A}(\nu)+2\pi\xi\bm{A}^\dagger(-\nu)-\sqrt{4\pi\Gamma}\bm{B}_\mathrm{out}(\nu).\label{eq2}
\end{align}

The solution to Eq.~\eqref{eq1} is
\begin{equation}
  \bm{A}(\nu)=\frac{\left[\Gamma-i\left(\nu+\delta)\right)\right]\sqrt{\Gamma/\pi}}{\left(\Gamma-i\nu\right)^2+\delta^2-\left\lvert\xi\right\rvert^2}\;\bm{b}_\nu+\frac{\xi\sqrt{\Gamma/\pi}}{\left(\Gamma-i\nu\right)^2+\delta^2-\left\lvert\xi\right\rvert^2}\;\bm{b}_{-\nu}^\dagger,
\end{equation}
and the solution to Eq.~\eqref{eq2} is
\begin{equation}
  \label{result}
  \bm{B}_\mathrm{out}(\nu)=\frac{\left(\Gamma-i\delta\right)^2+\nu^2+\left\lvert\xi\right\rvert^2}{\left(\Gamma-i\nu\right)^2+\delta^2-\left\lvert\xi\right\rvert^2}\;\bm{b}_\nu+\frac{2\Gamma\xi}{\left(\Gamma-i\nu\right)^2+\delta^2-\left\lvert\xi\right\rvert^2}\;\bm{b}_{-\nu}^\dagger.
\end{equation}

This is a Bogoliubov relation of the form
$\bm{B}_\mathrm{out}(\nu)=e^{i\phi_c}\cosh(\eta)\;\bm{b}_\nu+e^{i\phi_s}\sinh(\eta)\;\bm{b}_{-\nu}^\dagger$, with
\begin{equation}
    \phi_c(\nu,\Gamma,\lvert\xi\rvert,\delta)=\arctan\left[\frac{2\nu\Gamma}{\Gamma^2-\nu^2+\delta^2-\left\lvert\xi\right\rvert^2}\right]-\arctan\left[\frac{2\delta\Gamma}{\Gamma^2-\delta^2+\nu^2+\left\lvert\xi\right\rvert^2}\right];
\end{equation}
\begin{equation}
    \phi_s(\nu,\Gamma,\lvert\xi\rvert,\delta)=\arctan\left[\frac{2\nu\Gamma}{\Gamma^2-\nu^2+\delta^2-\left\lvert\xi\right\rvert^2}\right]+\arg\left[\xi\right];
\end{equation}
\begin{equation}
  \label{eta}
  \eta(\nu,\Gamma,\lvert\xi\rvert,\delta)=\frac{1}{2}\;\ln\left[\frac{\sqrt{(\Gamma^2+\nu^2+\left\lvert\xi\right\rvert^2-\delta^2)^2+4\delta^2\Gamma^2}+2\Gamma\left\lvert\xi\right\rvert}{\sqrt{(\Gamma^2+\nu^2+\left\lvert\xi\right\rvert^2-\delta^2)^2+4\delta^2\Gamma^2}-2\Gamma\left\lvert\xi\right\rvert}\right].
\end{equation}
\end{widetext}

On the `ridge', for $\delta=0$, this reduces to
\begin{equation}
    \eta(\nu,\Gamma,\lvert\xi\rvert)=\frac{1}{2}\;\ln\left[\frac{(\Gamma+\lvert\xi\rvert)^2+\nu^2}{(\Gamma-\lvert\xi\rvert)^2+\nu^2}\right].
\end{equation}

\section{Detection model}
At the output of the setup, we measure time-averaged statistics of the band-filtered single mode
\begin{equation}
  \label{eqB}
  \bm{B}_h(t)=\int_{-\infty}^{+\infty}\dd{\nu}h(\nu)\;\bm{B}_\mathrm{out}(\nu)\;e^{-i2\pi\nu t},
\end{equation}
where $h(\nu)$ is the integral-normalized ($\int \dd{\nu}\lvert h(\nu)\rvert^2$=1) filter response function of the detection filter. As $h(\nu)$ is integral-normalized, the quantity $\bm{N}_h(t)\equiv\bm{B}_h^\dagger(t)\bm{B}_h(t)$ is the number of photons counted per unit frequency and unit time. The statistical moments of the distribution, which are time-independent in the stationary regime, are
\begin{widetext}
\begin{multline}
    \mom{\bm{N}_h^k(t)}\equiv\mom{n^k}=\int_{-\infty}^{+\infty}\dd{\nu_1}\dd{\nu_2}\cdots\dd{\nu_{2k-1}}\dd{\nu_{2k}}\;\mom{\bm{B}_\mathrm{out}^\dagger(\nu_1)\bm{B}_\mathrm{out}(\nu_2)\cdots\bm{B}_\mathrm{out}^\dagger(\nu_{2k-1})\bm{B}_\mathrm{out}(\nu_{2k})}\\
    h^*(\nu_1)h(\nu_2)\cdots h^*(\nu_{2k-1})h(\nu_{2k})\,\exp[i(\nu_1-\nu_2+\cdots+\nu_{2k-1}-\nu_{2k})t].
\end{multline}
\end{widetext}
The quantum average is taken on the input mode (in our case, the
vacuum) after having substituted all $\bm{B}_\mathrm{out}$ by the appropriate
expressions of the free modes $\bm{b}_\nu$ using Eq.~\eqref{result}. Applying the usual commutation relations and the effect of the ladder operators on the vacuum, we find that the moments are indeed time-independent and verify
\begin{multline}
  \mom{n}=I_1;\quad\mom{\delta n^2}=I_1I_2+I_3;\\
  \mom{\delta n^3}=(I_1+I_2)(I_1I_2+3I_3),
\end{multline}
where
\begin{equation}
        I_1=\int_{-\infty}^{+\infty}\dd{\nu}\left\lvert h(\nu)\right\rvert^2 n(\nu),
\end{equation}
with $n(\nu)=\sinh^2[\eta(\nu)]$, and
\begin{equation}
    I_2=\int_{-\infty}^{+\infty}\dd{\nu}\left\lvert h(\nu)\right\rvert^2[n(\nu)+1];
\end{equation}
\begin{equation}
    I_3=\left\lvert\int_{-\infty}^{+\infty}\dd{\nu}h(\nu)h^*(-\nu)\sqrt{n(-\nu)[n(\nu)+1]}\right\rvert^2,
\end{equation}
with $\eta$ is defined in Eq.~\eqref{eta}.

We can make a few comments about these forms:
\begin{itemize}
    \item as $h(\nu)$ is integral-normalized, $I_2=I_1+1$;
    \item using Cauchy-Schwarz inequality, $I_3\le I_1I_2$, so that $\mom{\delta n^2}\le 2\mom{n}(\mom{n}+1)$.
\end{itemize}

Because of the exponential form of $\cosh$ and $\sinh$, it only takes $\eta$ to be of the order of a few units for the Cauchy-Schwarz inequality to become a quasi equality, so that we find, as expected for a squeezed vacuum,
\begin{equation}
    \label{JO}
    \begin{split}
        &\mom{\delta n^2}=2\mom{n}(\mom{n}+1);\\
        &\mom{\delta n^3}=2\mom{n}(2\mom{n}+1)(2\mom{n}+2).
    \end{split}
\end{equation}

In contrast, the statistics presented in~\cite{padurariu2012} corresponds to frequency-resolved photon detection. The measurement adds up the photocounts of independently-resolved orthogonal modes of width $\Delta\to 0$,
\begin{equation}
    \bm{B}_{\Delta,n}(t)=\int_{-\infty}^{+\infty}\dd{\nu}\bm{B}_\mathrm{out}(\nu)\gamma_{\Delta, n}(\nu)\;e^{-i2\pi\nu t},
\end{equation}
where we have defined
\begin{equation}
    \gamma_{\Delta, n}(\nu)\equiv\gamma_\Delta(\nu-n\Delta).
\end{equation}
Here, $\gamma_\Delta$ is a function of width $\Delta$, centered around $\nu=0$, verifying
\begin{equation}
    \int_{-\infty}^{+\infty}\dd{\nu}\gamma_\Delta(\nu-n\Delta)\gamma_\Delta(\nu-m\Delta)=\delta_{n,m}.
\end{equation}

Examples of mode generator functions are:
\begin{itemize}
    \item the sinc functions $\gamma_\Delta(\nu)=\sqrt{\frac{1}{\Delta}\mathrm{sinc}\left(\frac{\pi\,\nu}{\Delta}\right)}$;
    \item the window functions $\gamma_\Delta(\nu)=\frac{1}{\sqrt{\Delta}}\Pi_{\left[-\frac{\Delta}{2};\frac{\Delta}{2}\right]}(\nu)$.
\end{itemize}

In this theoretical framework, we find
\begin{equation}
    \begin{split}
        &\mom{n}_\Delta=\sum_{n} J_{1;n,n};\\
        &\mom{\delta n^2}_\Delta=\sum_{n,m} J_{1;n,m}J_{2;n,m}+J_{3;n,-m},
    \end{split}
\end{equation}
where
\begin{equation}
    J_{1;n,m}=\int_{-\infty}^{+\infty}\dd{\nu}\gamma_{n,\Delta}(\nu)\,\gamma_{m,\Delta}(\nu)\, n(\nu);
\end{equation}
\begin{equation}
    J_{2;n,m}=\int_{-\infty}^{+\infty}\dd{\nu}\gamma_{n,\Delta}(\nu)\,\gamma_{m,\Delta}(\nu)\,[n(\nu)+1];
\end{equation}
\begin{multline}
    J_{3;n,-m}= \left\lvert\int_{-\infty}^{+\infty}\dd{\nu}\gamma_{n,\Delta}(\nu)\,\gamma_{-m,\Delta}(\nu)\right.\\
    \left.\phantom{\int_{-\infty}^{+\infty}}\sqrt{n(-\nu)[n(\nu)+1]}\right\rvert^2.
\end{multline}

Using the orthonormality of the $\gamma_{n,\Delta}(\nu)$ functions, we have
\begin{multline}
    \int_{-\infty}^{+\infty}\dd{\nu}\gamma_\Delta(\nu-\nu_n)\,\gamma_\Delta(\nu-\nu_m)\;f(\nu)\simeq\delta_{n,m}\;f(\nu_n),
\end{multline}
and in the same limit of large $\eta(\nu)$ as in Eq.~\eqref{JO}, we obtain
\begin{equation}
    \mom{n}_\Delta=\sum_{n}N_n;\quad\mom{\delta n^2}_\Delta=\sum_{n}2N_{n}(N_{n}+1),
\end{equation}
with $N_{n}$ the number of photons in mode $n$.

\begin{figure}[ht]
    \centering
    \includegraphics[width=0.51\textwidth]{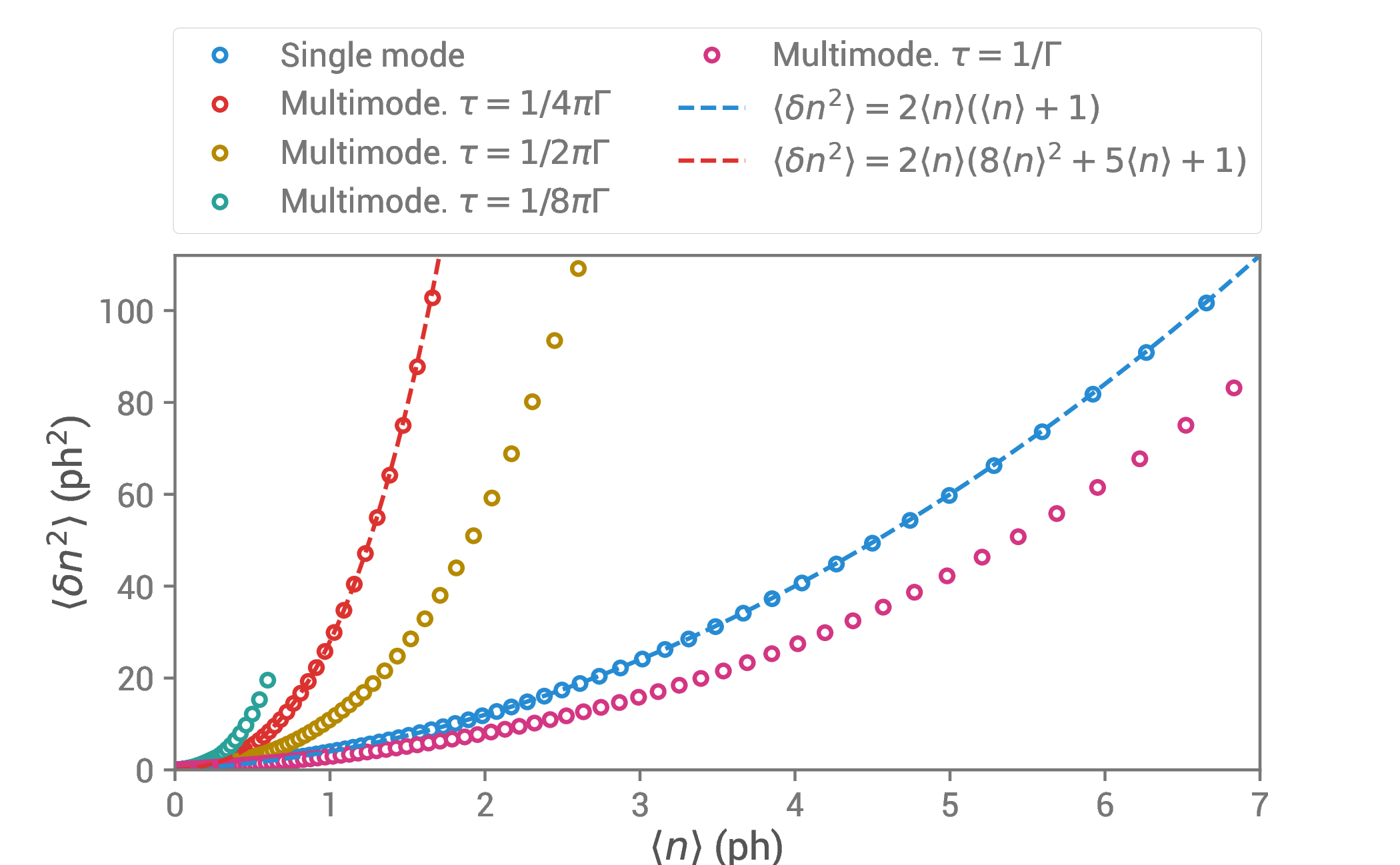}
    \caption{Comparison between single mode counting, as in the experiment presented in our paper, and multimode counting. In the latter case, the choice of integration time makes all the difference. For $\tau=1 /4\pi\Gamma$, we find the results of Ref.~\cite{padurariu2012}. However, when other multiples of $1 /\Gamma$ are chosen instead, the variance of photocounts behaves very differently and can actually be above or below that of the single mode case.}
    \label{figSV}
\end{figure}

We can compare these results with those of Eq.~\eqref{JO}, and it becomes apparent that we have a sum of independent modes, centered around their respective frequencies. As a result, we do not find the expected $\mom{\delta n^2}=2\mom{n}(\mom{n}+1)$, as there is an infinite number of modes.

Although always infinite because of the assumed infinite bandwidth, the number of modes increases linearly with the integration time $1 /\Delta$ required to resolve the modes. It is thus a good idea to define rates of photocount moments per unit time, $\Delta\mom{\delta n^k}_\Delta$, as we can expect them to remain finite. At the same time, we need to take into account the finite bandwidth of the detector, which translates in time domain to an integration time $\tau$ for the detection. The number of photons counted by the detector in any interval of time $\tau$ is thus $\tau\Delta\mom{\delta n^k}_\Delta$. Taking the limit of $\Delta\to 0$, we find
\begin{equation}
    \label{nDelta}
    \begin{split}
        \lim_{\Delta\to 0}\tau\Delta\mom{n}_\Delta&=\tau\int_{-\infty}^{+\infty}\dd{\nu}n(\nu),\\
        &=2\pi\Gamma\tau\, F_n(\Gamma\tau,\lvert\xi\rvert\tau,\delta\tau)
    \end{split}
\end{equation}
and
\begin{equation}
    \label{dn2Delta}
    \begin{split}
        \lim_{\Delta\to 0}\tau\Delta\mom{\delta n^2}_\Delta&=\tau\int_{-\infty}^{+\infty}\dd{\nu}2\,n(\nu)[n(\nu)+1].\\
        &=2\pi\Gamma\tau\, F_{\delta n^2}(\Gamma\tau,\lvert\xi\rvert\tau,\delta\tau),
    \end{split}
\end{equation}
where $F_n$ and $F_{\delta n^2}$ are unitless functions of the unitless parameters $\Gamma\tau, \abs{\xi}\tau, \delta\tau$. In that sense, they are universal and useful for the characterization of the noise of paramps.

The Fano factor is thus simply
\begin{equation}
    \mathcal{F}=\frac{F_{\delta n^2}(\Gamma\tau,\lvert\xi\rvert\tau,\delta\tau)}{F_n(\Gamma\tau,\lvert\xi\rvert\tau,\delta\tau)},
\end{equation}
and it depends on the valuse of the unitless parameters $\Gamma\tau, \abs{\xi}\tau, \delta\tau$.

The Fano factor found in Ref.~\cite{padurariu2012}, corresponds to defining $\tau\equiv1 /4\pi\Gamma$, $\delta=0$, and using the free parameter $\abs{\xi}\tau$ (proportional to the pump power) as a knob to change the average photon number in the cavity. In this specific context we find
\begin{equation}
    \label{Nazarovok}
    \mom{\delta n^2}=2\mom{n}\left[8\mom{n}^2+5\mom{n}+1\right],
\end{equation}
as illustrated in Fig.~\ref{figSV}.

If we instead choose $\tau=1 /2\pi\Gamma$, $\tau=1 /8\pi\Gamma$, or even $\tau=1 /\Gamma$ as the integration time, we find different relations between $\mom{\delta n^2}$ and $\mom{n}$, also shown in Fig.~\ref{figSV}.

\section{Wideband detection}
The quantity that is detected in our experiments is voltage, the equivalent of the electric field in free space. As shown in Ref.~\cite{Virally2019}, the voltage is not the same thing as the `photonic field' carrying information about photocounts on a detector. So what is really measured is not directly the bosonic mode of Eq.~\eqref{eqB}, but rather the linked quantity,
\begin{equation}
  \label{eqV}
  \bm{V}_h(t)=\sqrt{\nu_0}\int_{-\infty}^{+\infty}\dd{\nu}\sqrt{1+\frac{\nu}{\nu_0}}\;h(\nu)\;\bm{B}_\mathrm{out}(\nu)\;e^{-i2\pi\nu t},
\end{equation}
where $\nu_0$ is the central frequency (6~GHz) of the detection band.

\bibliography{article.bib}